\documentclass[runningheads]{llncs}
\usepackage{graphicx}

\usepackage{amssymb}
\usepackage{amsmath}

\usepackage{hyperref}

\begin{document}


\title{Computationally Efficient 3D MRI Reconstruction with Adaptive MLP}

\author{Eric Z. Chen\inst{1}, Chi Zhang\inst{2}\thanks{Contribution from Chi Zhang was carried out during his internship at United Imaging Intelligence, Cambridge, MA.}, Xiao Chen\inst{1}, Yikang Liu\inst{1}, Terrence Chen\inst{1}, Shanhui Sun\inst{1}}

\authorrunning{Chen et al.}
\institute{United Imaging Intelligence, Cambridge, MA, USA \and Department of Electrical and Computer Engineering, and Center for Magnetic Resonance Research, University of Minnesota, Minneapolis, MN, USA }


\maketitle

\begin{abstract}
Compared with 2D MRI, 3D MRI provides superior volumetric spatial resolution and signal-to-noise ratio. However, it is more challenging to reconstruct 3D MRI images. Current methods are mainly based on convolutional neural networks (CNN) with small kernels, which are difficult to scale up to have sufficient fitting power for 3D MRI reconstruction due to the large image size and GPU memory constraint. Furthermore, MRI reconstruction is a deconvolution problem, which demands long-distance information that is difficult to capture by CNNs with small convolution kernels. The multi-layer perceptron (MLP) can model such long-distance information, but it requires a fixed input size. In this paper, we proposed Recon3DMLP,  a hybrid of CNN modules with small kernels for low-frequency reconstruction and adaptive MLP (dMLP) modules with large kernels to boost the high-frequency reconstruction, for 3D MRI reconstruction. We further utilized the circular shift operation based on MRI physics such that dMLP accepts arbitrary image size and can extract global information from the entire FOV. We also propose a GPU memory efficient data fidelity module that can reduce $>$50$\%$ memory. We compared Recon3DMLP with other CNN-based models on a high-resolution (HR) 3D MRI dataset. Recon3DMLP improves HR 3D reconstruction and outperforms several existing CNN-based models under similar GPU memory consumption, which demonstrates that Recon3DMLP is a practical solution for HR 3D MRI reconstruction.

\end{abstract}

\section{Introduction}

Compared with 2D MRI, 3D MRI has superior volumetric spatial resolution and signal-to-noise ratio. However, 3D MRI, especially high resolution (HR) 3D MRI (e.g., at least $1mm^3$ voxel size), often takes much longer acquisition time than 2D scans. Therefore, it is necessary to accelerate 3D MRI by acquiring sub-sampled k-space. However, it is more challenging to reconstruct HR 3D MRI images than 2D images. For example, HR 3D MRI data can be as large as 380$\times$294$\times$138$\times$64, which is more than 100X larger than common 2D MRI data \cite{fastMRI}(e.g., 320$\times$320$\times$1$\times$15, hereafter data dimensions are defined as RO$\times$PE$\times$SPE$\times$Coil, where RO stands for read-out, PE for phase-encoding, and SPE for slice-phase-encoding). Although deep learning (DL) based methods have shown superior reconstruction speed and image quality, they are constrained by GPU memory for 3D MRI reconstruction in the clinical setting. 

Due to the large 3D image size and computation constraint, the state-of-the-art methods for 2D MRI reconstruction \cite{knoll2020advancing,muckley2021results} are not directly transferable to 3D MRI reconstruction. 
Instead of using 3D convolutions,  \cite{ahn2022deep} proposed a 2D CNN on the PE-SPE plane for 3D MRI reconstruction. \cite{R4} proposed to downsample the 3D volume and reconstruct the smaller 3D image, which is then restored to the original resolution by a super-resolution network. \cite{ericpgdl,R7} used 3D CNN models to reconstruct each coil of 3D MRI data independently. \cite{MEL} applied the gradient checkpointing technique to save the GPU memory during training. GLEAM \cite{R6} splits the network into modules and updates the gradient on each module independently, which reduces memory usage during training.
 
\begin{figure}[!t]
\centering
\includegraphics[scale=0.25]{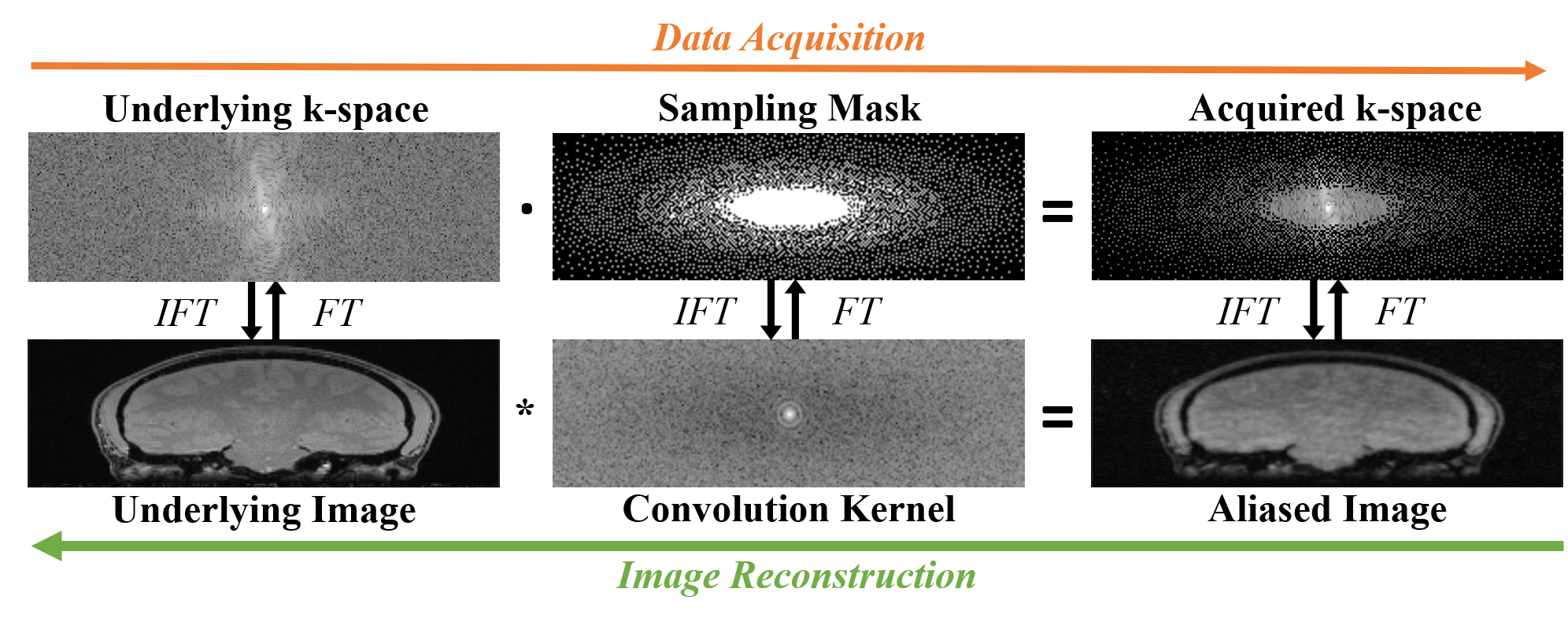}
\caption{Demonstration of k-space acquisition, which is equivalent to a convolution in the image domain, and reconstruction, which is a deconvolution process to recover the underlying image. The convolution kernel has the most energy at the center but spans the entire FOV, suggesting that global information is necessary for reconstruction.}
\label{fig:acceleratedMRI}
\end{figure}

The previous works on 3D MRI reconstruction have several limitations. 
First, all these methods are based on CNN. In the context of 3D reconstruction, deep CNN networks require significant GPU memory and are difficult to scale. As a result, many models are designed to be relatively small to fit within available resources \cite{ahn2022deep,ericpgdl,R7}. Given that a high-resolution 3D volume can contain over 100 million voxels, the model's fitting power is critical. Small models may lack the necessary fitting power, resulting in suboptimal performance in 3D MRI reconstruction. 
Second, due to network inductive bias, CNN prioritizes low-frequency information reconstruction and tends to generate smooth images \cite{CNNfreqBias,CNNfreqBias2}. 
Third, CNN has a limited receptive field due to highly localized convolutions using small kernels. The k-space sub-sampling is equivalent to convolving the underlying aliasing-free image using a kernel that covers the entire field of view (FOV) (orange arrow in Fig. \ref{fig:acceleratedMRI}). Therefore, the contribution of aliasing artifacts for a voxel comes from all other voxels globally in the sub-sampling directions. Then reconstruction is deconvolution and it is desirable to utilize the global information along the sub-sampled directions (green arrow in Fig. \ref{fig:acceleratedMRI}).
Although convolution-based methods such as large kernels \cite{51Conv,DeConvCNN}, dilation, deformable convolution \cite{DeformConv} as well as attention-based methods such as Transformers \cite{ViT,SwinViT} can enlarge the receptive field, it either only utilizes limited voxels within the FOV or may lead to massive computation \cite{CNNfreqBias}.
Recently, multi-layer perceptron (MLP) based models have been proposed for various computer vision tasks \cite{MLPMIXER,MAXIM,lian2021mlp,CycleMLP,ConvMLP,ResMLP,MLPMRIDenoise,valanarasu2022unext}. MLP models have better fitting power and less inductive bias than CNN models \cite{MLPsurvey}. MLP performs matrix multiplication instead of convolution, leading to enlarged receptive fields with lower memory and time cost than CNN and attention-based methods. However, MLP requires a fixed input image resolution and several solutions have been proposed \cite{CycleMLP,lian2021mlp,SwinViT,MLPsurvey}. Nevertheless, these methods were proposed for natural image processing and failed to exploit global information from the entire FOV. Img2ImgMixer\cite{mansour2022image} adapted MLP-Mixer \cite{MLPMIXER} to 2D MRI reconstruction but on fixed-size images. AUTOMAP \cite{AUTOMAP} employs MLP on whole k-space to learn the Fourier transform, which requires massive GPU memory and a fixed input size and thus is impractical even for 2D MRI reconstruction. 
Fourth, the methods to reduce GPU memory are designed to optimize gradient calculation for training, which is not beneficial for inference when deployed in clinical practice.

To tackle these problems, we proposed Recon3DMLP for 3D MRI reconstruction, a hybrid of CNN modules with small kernels for low-frequency reconstruction and adaptive MLP (dMLP) modules with large kernels to boost the high-frequency reconstruction. The dMLP improves the model fitting ability with almost the same GPU memory usage and a minor increase in computation time. We utilized the circular shift operation \cite{SwinViT} based on MRI physics such that the proposed dMLP accepts arbitrary image size and can extract global information from the entire FOV. Furthermore, we propose a memory-efficient data fidelity (eDF) module that can reduce $>$50$\%$ memory. We also applied gradient checkpointing, RO cropping, and half-precision (FP16) to save GPU memory. We compared Recon3DMLP with other CNN-based models on an HR 3D multi-coil MRI dataset. The proposed dMLP improves HR 3D reconstruction and outperforms several existing CNN-based strategies under similar GPU memory consumption, which demonstrate that Recon3DMLP is a practical solution for HR 3D MRI reconstruction.

\section{Method}

\subsection{Recon3DMLP for 3D MRI Reconstruction}
The MRI reconstruction problem can be solved as
\begin{equation}\label{eq:DFstep}
    x = \arg \min_{x} ||{{y}-{MFSx}}||_2^2 + \lambda ||{x-g_{\theta}(x_u)}||_2^2,
\end{equation}
where $y$ is the acquired measurements, $x_u$ is the under-sampled image, $M$ and $S$ are the sampling mask and coil sensitivities, $F$ denotes FFT and $\lambda$ is a weighting scalar. $g_{\theta}$ is a neural network with the data fidelity (DF) module \cite{TheCascade}.

The proposed Recon3DMLP adapts the memory-friendly cascaded structure. Previous work has shown that convolutions with small kernels are essential for low-level tasks \cite{MAXIM}. Therefore, we added the dMLP module with large kernels after each 3D CNN with small kernels (k=3) to increase the fitting capacity and utilize the global information.

\begin{figure}[!ht]
\centering
\includegraphics[scale=0.18]{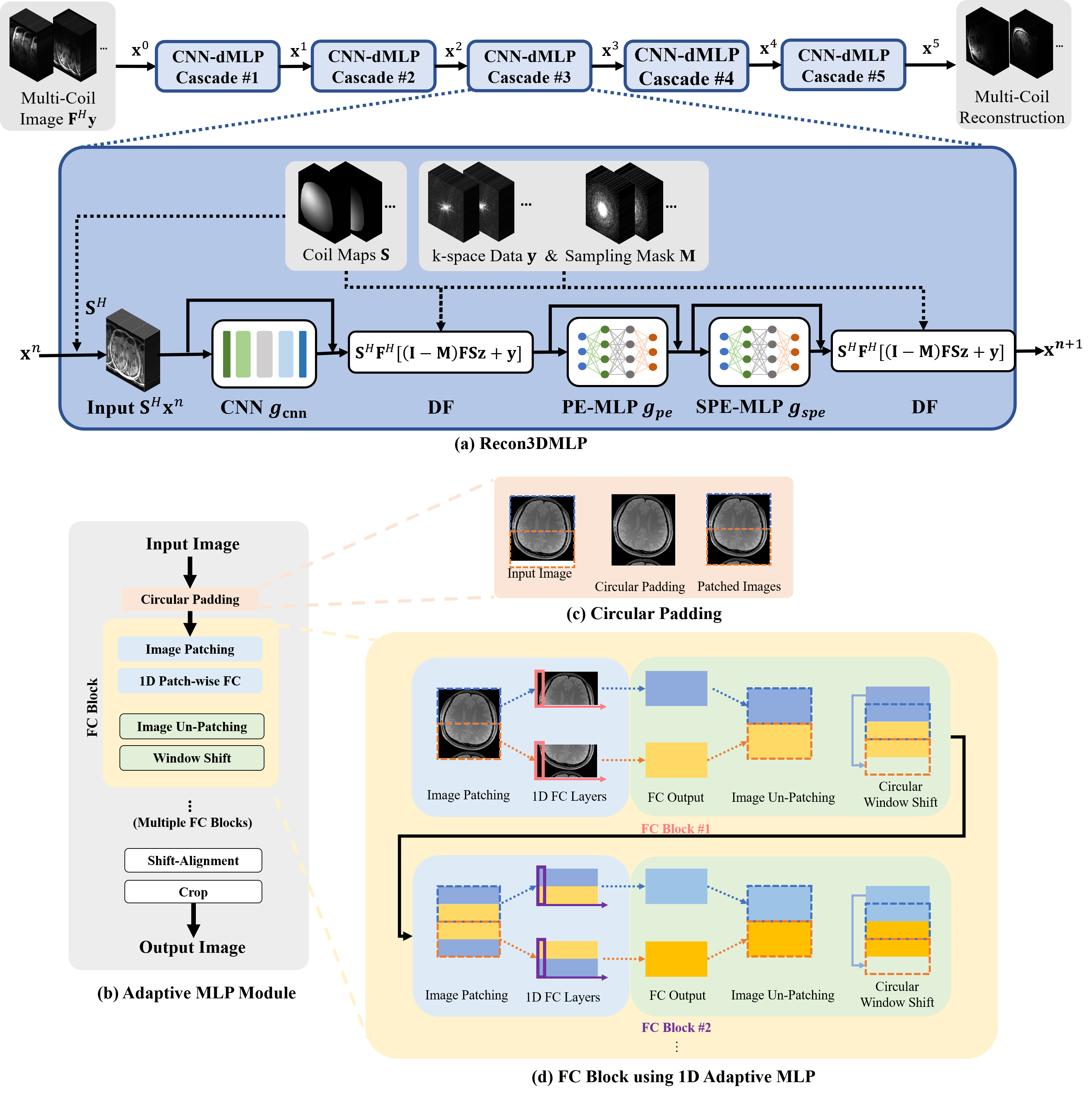}
\caption{(a) The proposed Recon3DMLP for 3D MRI reconstruction, which is a cascaded network and each cascade consists of a hybrid of CNN and dMLP modules. (b) The overall structure of dMLP module. (c) Circular padding is applied to ensure image can be patched. (d) Shared 1D FC layers is then applied to the patch dimension, followed by un-patch and shift operations. The FC blocks are stacked multiple times. The shift-alignment and crop operations are then applied to recover the original image shape. }
\label{fig:themodel}
\end{figure}

\subsection{Adaptive MLP for Flexible Image Resolution}

The dMLP module includes the following operations (Fig.\ref{fig:themodel}): 1) circular padding, 2) image patching, 3) FC layers, 4) circular shift, 5) shift alignment and 6) cropping. The input is circular-padded in order to be cropped into patches, and the shared 1D FC layers are applied over the patch dimension. The output is then un-patched into the original image shape. Next, the circular shift is applied along the patched dimension by a step size. The circular padding and shift are based on the DFT periodicity property of images. Then operations 2-4 (FC block) are stacked several times. Due to the shift operation in each FC block, the current patch contains a portion of information from two adjacent patches in the previous FC block, which allows information exchange between patches and thus dMLP can cover the entire FOV. In the end, the shift alignment is applied to roll back the previous shifts in the image domain. The padded region is then cropped out to generate the final output. Since the sub-sampling in k-space is a linear process that can be decomposed as 1D convolutions in the image domain along each sub-sampled direction, we use 1D dMLP for 3D reconstruction.

\subsection{Memory Efficient Data Fidelity Module}
In the naive implementation of the DF module
\begin{equation}
d_{DF} = {S}^H {F}^H[(I - M) FSz+y],
\end{equation}
the coil combined image $z$ is broadcasted to multi-coil data $(I - M)FSz$ and it increases memory consumption. Instead, we can process the data coil-by-coil
\begin{equation}\label{eq:compactDF2}
d_{eDF} 
                = \sum_{c}{S}_c^H {F}^H[{(I - M_c) FS}_cz+y_c],
\end{equation}
where $c$ is the coil index. Together with eDF,  we also employed RO cropping and gradient checkpointing for training and half-precision for inference.

\subsection{Experiments}
We collected a multi-contrast HR 3D brain MRI dataset with IRB approval, ranging from 224$\times$220$\times$96$\times$12 to 336$\times$336$\times$192$\times$32 \cite{ericpgdl,ye2022multi}. There are 751 3D multi-coil images for training, 32 for validation, and 29 for testing. 

We started with a small 3D CNN model (Recon3DCNN) with an expansion factor $e=6$, where the channels increase from 2 to 12 in the first convolution layer and reduce to 2 in the last layer in each cascade. We then enlarged Recon3DCNN with increased width (e=6,12,16,24) and depth (double convolution layers in each cascade). We also replaced the 3D convolution in Recon3DCNN with depth separable convolution \cite{howard2017mobilenets} or separate 1D convolution for each 3D dimension. We also adapted the reparameterization technique \cite{ding2021repvgg} for Recon3DCNN such that the residual connection can be removed during inference to reduce the GPU memory. For comparison, we also adapted a 3D version of cascaded UNet, where each UNet has five levels with e=4 at the initial layer and the channels were doubled at each level. 
To demonstrate the effectiveness of dMLP, we built Recon3DMLP by adding two 1D dMLP on PE (k=64) and SPE (k=16) to the smallest Recon3DCNN (e=6). Since GELU \cite{hendrycks2016gaussian,MLPMIXER} has larger memory overhead, we used leaky ReLU for all models. 
We performed ablation studies on Recon3DMLP by sharing the FC blocks among shifts, removing shifts, reducing patch size to 3 as well as replacing the dMLP with large kernel convolutions (LKconv) using k=65 for PE and k=17 for SPE, as well as small kernel convolutions (SKconv) using k=3. 
We attempted to adapt ReconFormer \footnote{\url{https://github.com/guopengf/ReconFormer}}, a transformer-based model, and Img2ImgMixer \footnote{\url{https://github.com/MLI-lab/imaging_MLPs}}, an MLP based model. Both models require to specify a fixed input image size when constructing the model and failed to run on datasets with various sizes, indicating the limitation of these methods. Note that the two models were originally demonstrated on the 2D datasets with the same size \cite{guo2022reconformer,mansour2022image}. All models were trained with loss=L1+SSIM and lr=0.001 for 50 epochs using an NVIDIA A100 GPU with Pytorch 1.10 and CUDA 11.3. The pvalues were calculated by the Wilcoxon signed-ranks test.

\begin{figure}[t]
\centering
\includegraphics[scale=0.38]{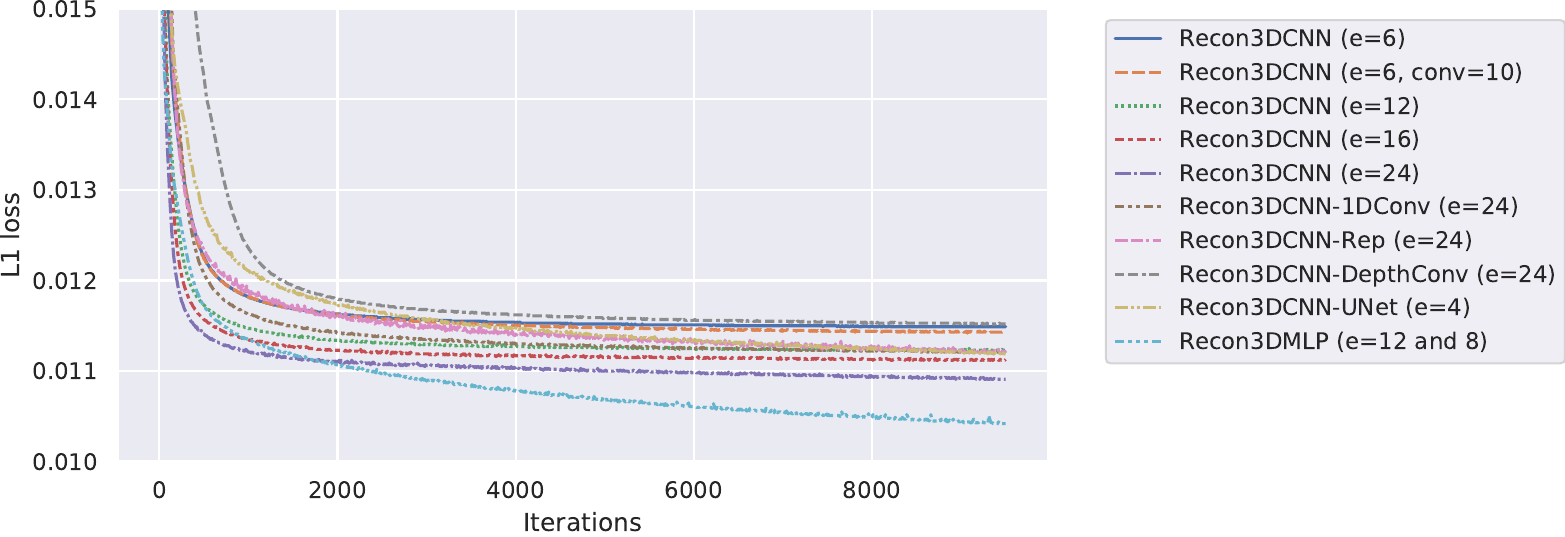}
\caption{The fitting power of various models on HR 3D MRI reconstruction. Models with lower loss indicate better fitting capacity. } 
\label{fig:overfitting}
\end{figure}


\begin{table}[t]
\tiny
\centering
\caption{Evaluation of different models on HR 3D MRI reconstruction. The inference GPU memory and forward time were measured on a 3D image in 380$\times$294$\times$138$\times$24.}\label{table:metrics}
\begin{tabular}{|l|r|r|r|r|r|r|}
\hline
Model & Memory Saving  & Parameters (K)  & GPU (G)& Time (S) & SSIM  & PSNR  \\
\hline
Recon3DCNN (e=6) & None & 65  & {\textgreater{}40} & {Fail} & {NA} & {NA}   \\
\hline
Recon3DCNN (e=6) &  FP16  & 65  & 35.5               & 1.17 & 0.9581 & 40.2790  \\

Recon3DCNN (e=6)  & eDF  & 65  & 18.8               & 3.49 & 0.9581 & 40.2795  \\ 
Recon3DCNN (e=6) & FP16+eDF & 65 & 11.5 & 3.04 & 0.9581 & 40.2785   \\
\hline
Recon3DCNN (e=6, conv=10)  & FP16+eDF  & 130  & 11.5  & 4.26   & 0.9597  & 40.5042    \\
Recon3DCNN (e=12) & FP16+eDF & 247  & 11.6 & 3.14 & 0.9623& 40.8118 \\
Recon3DCNN (e=16)  & FP16+eDF & 433  & 13.3  & 3.20 & 0.9636 & 40.9880  \\
Recon3DCNN (e=24) & FP16+eDF & 960 &  15.2 &  3.67 &  0.9649 &  41.1503  \\
\hline
Recon3DCNN-1DConv (e=24) & FP16+eDF & 386  & 16.6  & 5.51   & 0.9639  & 41.0473  \\
Recon3DCNN-Rep (e=24) & FP16+eDF  & 995   & 12.5   & 3.68   & 0.9613 & 40.4970    \\
Recon3DCNN-DepthConv (e=24)  & FP16+eDF & 111  & 17.2  & 4.11 & 0.9594  & 40.4367 \\		
Recon3DCNN-UNet (e=4) & FP16+eDF & 7,056  & 10.6 & 4.16  & 0.9565  & 40.4229  \\
\hline
Recon3DMLP (e=6/8, SKconv) & FP16+eDF  & 72 & 10.5 & 4.38  & 0.9617 & 41.0456   \\
Recon3DMLP (e=6/8, LKconv) & FP16+eDF   & 157  & 11.5   & 4.55  & 0.9620   & 41.0741   \\ 
Recon3DMLP (e=6/8, k=3) & FP16+eDF   & 115   & 11.5   & 3.37 & 0.9622    & 41.0627\\  
Recon3DMLP (e=6/8, share) & FP16+eDF & 1,465 & 11.5 & 3.36 & 0.9627 & 41.1455          \\
Recon3DMLP (e=6/8, no shift) & FP16+eDF & 11,264 & 11.5  & 3.36 & 0.9619  & 41.0853          \\
Recon3DMLP (e=6/8, proposed) & FP16+eDF  & 11,264 &  11.5 &  3.38  & 0.9637 &  41.1953  \\
\hline
\end{tabular}
\end{table}

\section{Results}

We first demonstrate the benefit of eDF and FP16 inference with a small CNN model Recon3DCNN (e=6) (first and second panels in Table \ref{table:metrics}). Without eDF and FP16, the model takes $>$40G inference GPU memory and fails to reconstruct the test data, which indicates the challenge of HR 3D MRI reconstruction. FP16 and eDF reduce at least 11\% and 53\% inference memory. However, the model with only eDF is slower than the model with only FP16. By combining eDF and FP16, the inference GPU memory is reduced by 71\% to 11.5G, which makes the model feasible to be deployed with a mid-range GPU in practice. Hereafter, we applied eDF and FP16 to all models. 

\begin{figure}[!ht]
\centering
\includegraphics[scale=0.48]{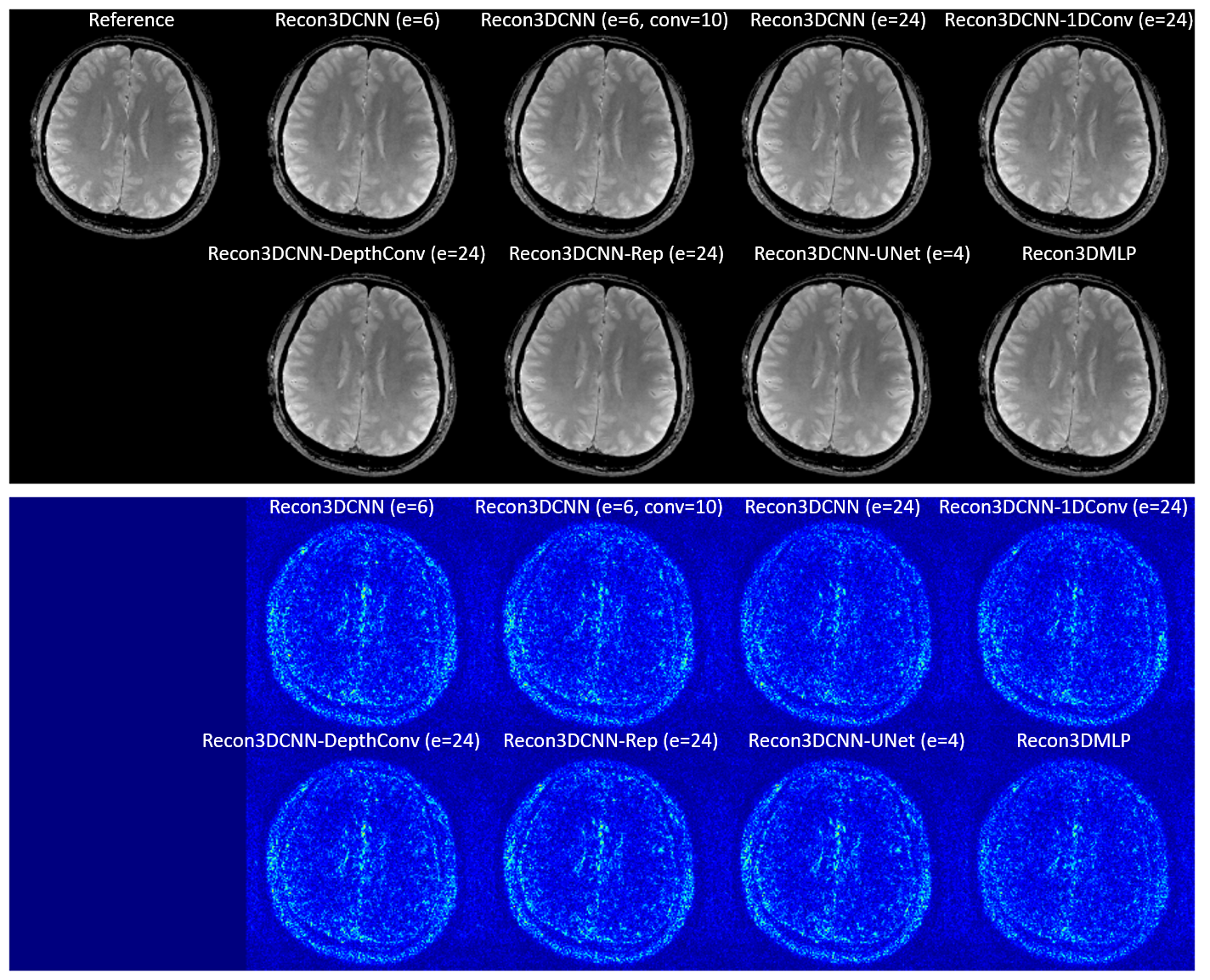}
\caption{Reconstruction results and corresponding error maps. 
}
\label{fig:result}
\end{figure}

Next, we aim to improve Recon3DCNN's performance by increasing the width and depth (third panel in Table \ref{table:metrics} and Fig. \ref{fig:result}). By making the model wider (increase e=6 to e=24), the PSNR/SSIM improves significantly ($p<10^{-7}$). However, the inference GPU memory also increases by 33\%. On the other hand, doubling the depth also improves the performance ($p<10^{-5}$), but not as significantly as increasing the model width. Also, the former increases inference time (40\%) more than the latter (21\%). Also increasing the model depth does not affect the inference GPU memory. Next, we experimented with those commonly used techniques for efficient computation to modify the best CNN model Recon3DCNN (e=24) (fourth panel in Table \ref{table:metrics} and Fig. \ref{fig:result}). All those variants lead to a performance drop compared to the original model ($p<10^{-7}$), because such methods reduce the model's fitting capacity. Those variants also result in memory increase except Recon3DCNN with reparameterization technique. These results indicate such methods proposed for natural image processing are not suitable for HR 3D MRI reconstruction.

The performance of Recon3DCNN improves when becoming larger (i.e., more parameters), which indicates CNN models lack fitting power for HR 3D MR reconstruction. Therefore, we performed an overfitting experiment where models were trained and tested on one data. Fig. \ref{fig:overfitting} confirms that Recon3DCNN can not overfit one test data in 10K iterations and models with better fitting ability tend to have better PSNR/SSIM (Table \ref{table:metrics})). The variants of Recon3DCNN indeed have lower fitting power than the original model. This motivates us to build Recon3DMLP by adding dMLP to Recon3DCNN (e=6) to increase its capacity while maintaining low memory usage. Recon3DMLP has better fitting ability and less reconstruction error than all models (Fig. \ref{fig:overfitting} and \ref{fig:result}). Compared to the smaller Recon3DCNN (e=6),  Recon3DMLP has similar GPU memory usage but better PSNR/SSIM ($p<10^{-7}$).   Compared to the larger Recon3DCNN (e=24),  Recon3DMLP has 24\% less GPU memory usage and better PSNR ($p<10^{-7}$) and only marginally worse SSIM ($p=0.05$). The cascaded 3D UNet has less GPU memory consumption but lower fitting power, worse performance ($p<10^{-7}$) and longer inference time than Recon3DCNN (e=24) and Recon3DMLP.

\begin{figure}[!ht]
\centering
\includegraphics[width=\textwidth]{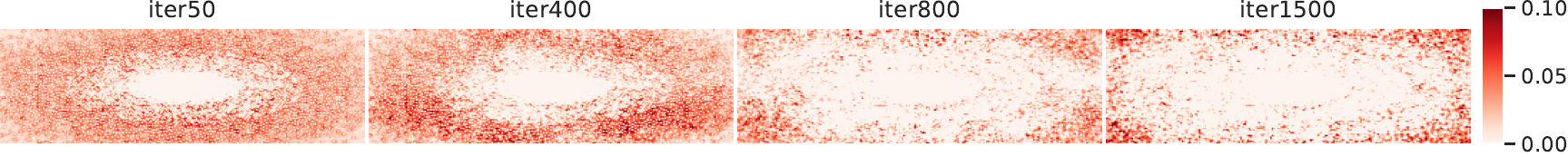}
\caption{The k-space difference between Recon3DMLP with and without dMLP across training iterations. Red areas in the outer k-space indicate Recon3DMLP with dMLP has recovered more high-frequency information and faster than that without dMLP.}
\label{fig:kspace_difference}
\end{figure}

To investigate the source of the gain, we perform ablation studies on Recon3DMLP (last panel in Table \ref{table:metrics}). By removing the shift operations, the dMLP module can only utilize the global information within the large patch, which leads to a drop in PSNR/SSIM ($p<10^{-7}$). When reducing the patch size to 3 but keeping the shift operations such that the model can only utilize the global information through the shift operations, the performance also drops ($p<10^{-7}$) but less than the previous one. This indicates the shift operations can help the model to learn the global information and thus improve the reconstruction results. Also, models with and without shift operations do not significantly differ in GPU memory and time, suggesting the shift operations are computationally efficient. By sharing the FC parameters among shifts, the model has much fewer parameters and performance drops slightly ($p<10^{-7}$) while GPU memory and time are similar to the original Recon3DMLP. We also replaced the dMLP modules in Recon3DMLP with convolutions using larger kernels and small kernels, respectively. Recon3DMLP (LKconv) and Recon3DMLP (SKconv) \footnote{These are CNN models but we consider them as ablated models of Recon3DMLP and slightly abuse the notation.} have worse performance ($p<10^{-3}$) as well as longer time than their counterpart Recon3DMLP and Recon3DMLP (small patch), indicating the dMLP is better than the convolutions for HR 3D MRI reconstruction. We compared the Recon3DMLP with and without dMLP modules and Fig. \ref{fig:kspace_difference} shows that dMLP modules help to learn the high-frequency information faster.

\section{Discussion and Conclusion}

Although MLP has been proposed for vision tasks on natural images as well as 2D MRI reconstruction with fixed input size, we are the first to present a practical solution utilizing the proposed dMLP and eDF to overcome the computational constraint for HR 3D MRI reconstruction with various sizes. Compared with CNN based models, Recon3DMLP improves image quality with a little increase in computation time and similar GPU memory usage.

One limitation of our work is using the same shift and patch size without utilizing the multi-scale information. dMLP module that utilizes various patch and shift sizes will be investigated in future work. MLP-based models such as Recon3DMLP may fail if the training data is small.

\bibliographystyle{splncs04}
\bibliography{refs}

\begin{thebibliography}{10}
\providecommand{\url}[1]{\texttt{#1}}
\providecommand{\urlprefix}{URL }
\providecommand{\doi}[1]{https://doi.org/#1}

\bibitem{ahn2022deep}
Ahn, S., Wollner, U., McKinnon, G., Jansen, I.H., Brada, R., Rettmann, D.,
  Cashen, T.A., Huston, J., DeMarco, J.K., Shih, R.Y., et~al.: Deep
  learning-based reconstruction of highly accelerated 3d mri. arXiv preprint
  arXiv:2203.04674  (2022)

\bibitem{CNNfreqBias2}
Basri, R., Galun, M., Geifman, A., Jacobs, D., Kasten, Y., Kritchman, S.:
  Frequency bias in neural networks for input of non-uniform density. In:
  International Conference on Machine Learning. pp. 685--694. PMLR (2020)

\bibitem{ericpgdl}
Chen, E.Z., Ye, Y., Chen, X., Lyu, J., Zhang, Z., Hu, Y., Chen, T., Xu, J.,
  Sun, S.: Accelerating 3d multiplex mri reconstruction with deep learning.
  arXiv preprint arXiv:2105.08163  (2021)

\bibitem{CycleMLP}
Chen, S., Xie, E., Ge, C., Liang, D., Luo, P.: Cyclemlp: A mlp-like
  architecture for dense prediction. arXiv preprint arXiv:2107.10224  (2021)

\bibitem{DeformConv}
Dai, J., Qi, H., Xiong, Y., Li, Y., Zhang, G., Hu, H., Wei, Y.: Deformable
  convolutional networks. In: Proceedings of the IEEE international conference
  on computer vision. pp. 764--773 (2017)

\bibitem{ding2021repvgg}
Ding, X., Zhang, X., Ma, N., Han, J., Ding, G., Sun, J.: Repvgg: Making
  vgg-style convnets great again. In: Proceedings of the IEEE/CVF conference on
  computer vision and pattern recognition. pp. 13733--13742 (2021)

\bibitem{ViT}
Dosovitskiy, A., Beyer, L., Kolesnikov, A., Weissenborn, D., Zhai, X.,
  Unterthiner, T., Dehghani, M., Minderer, M., Heigold, G., Gelly, S., et~al.:
  An image is worth 16x16 words: Transformers for image recognition at scale.
  arXiv preprint arXiv:2010.11929  (2020)

\bibitem{guo2022reconformer}
Guo, P., Mei, Y., Zhou, J., Jiang, S., Patel, V.M.: Reconformer: Accelerated
  mri reconstruction using recurrent transformer. arXiv preprint
  arXiv:2201.09376  (2022)

\bibitem{hendrycks2016gaussian}
Hendrycks, D., Gimpel, K.: Gaussian error linear units (gelus). arXiv preprint
  arXiv:1606.08415  (2016)

\bibitem{howard2017mobilenets}
Howard, A.G., Zhu, M., Chen, B., Kalenichenko, D., Wang, W., Weyand, T.,
  Andreetto, M., Adam, H.: Mobilenets: Efficient convolutional neural networks
  for mobile vision applications. arXiv preprint arXiv:1704.04861  (2017)

\bibitem{MEL}
Kellman, M., Zhang, K., Markley, E., Tamir, J., Bostan, E., Lustig, M., Waller,
  L.: Memory-efficient learning for large-scale computational imaging. IEEE
  IEEE Trans Comp Imag  \textbf{6},  1403--1414 (2020)

\bibitem{knoll2020advancing}
Knoll, F., Murrell, T., Sriram, A., Yakubova, N., Zbontar, J., Rabbat, M.,
  Defazio, A., Muckley, M.J., Sodickson, D.K., Zitnick, C.L., et~al.: Advancing
  machine learning for mr image reconstruction with an open competition:
  Overview of the 2019 fastmri challenge. Magnetic resonance in medicine
  \textbf{84}(6),  3054--3070 (2020)

\bibitem{fastMRI}
Knoll, F., Zbontar, J., Sriram, A., Muckley, M.J., Bruno, M., Defazio, A.,
  Parente, M., Geras, K.J., Katsnelson, J., Chandarana, H., et~al.: fastmri: A
  publicly available raw k-space and dicom dataset of knee images for
  accelerated mr image reconstruction using machine learning. Radiology:
  Artificial Intelligence  \textbf{2}(1),  e190007 (2020)

\bibitem{ConvMLP}
Li, J., Hassani, A., Walton, S., Shi, H.: Convmlp: Hierarchical convolutional
  mlps for vision. arXiv preprint arXiv:2109.04454  (2021)

\bibitem{lian2021mlp}
Lian, D., Yu, Z., Sun, X., Gao, S.: As-mlp: An axial shifted mlp architecture
  for vision. arXiv preprint arXiv:2107.08391  (2021)

\bibitem{MLPsurvey}
Liu, R., Li, Y., Tao, L., Liang, D., Zheng, H.T.: Are we ready for a new
  paradigm shift? a survey on visual deep mlp. Patterns  \textbf{3}(7),  100520
  (2022)

\bibitem{51Conv}
Liu, S., Chen, T., Chen, X., Chen, X., Xiao, Q., Wu, B., Pechenizkiy, M.,
  Mocanu, D., Wang, Z.: More convnets in the 2020s: Scaling up kernels beyond
  51x51 using sparsity. arXiv preprint arXiv:2207.03620  (2022)

\bibitem{SwinViT}
Liu, Z., Lin, Y., Cao, Y., Hu, H., Wei, Y., Zhang, Z., Lin, S., Guo, B.: Swin
  transformer: Hierarchical vision transformer using shifted windows. In:
  Proceedings of the IEEE/CVF international conference on computer vision. pp.
  10012--10022 (2021)

\bibitem{mansour2022image}
Mansour, Y., Lin, K., Heckel, R.: Image-to-image mlp-mixer for image
  reconstruction. arXiv preprint arXiv:2202.02018  (2022)

\bibitem{muckley2021results}
Muckley, M.J., Riemenschneider, B., Radmanesh, A., Kim, S., Jeong, G., Ko, J.,
  Jun, Y., Shin, H., Hwang, D., Mostapha, M., et~al.: Results of the 2020
  fastmri challenge for machine learning mr image reconstruction. IEEE
  transactions on medical imaging  \textbf{40}(9),  2306--2317 (2021)

\bibitem{R6}
Ozturkler, B., Sahiner, A., Ergen, T., Desai, A.D., Sandino, C.M., Vasanawala,
  S., Pauly, J.M., Mardani, M., Pilanci, M.: Gleam: Greedy learning for
  large-scale accelerated mri reconstruction. arXiv preprint arXiv:2207.08393
  (2022)

\bibitem{CNNfreqBias}
Rahaman, N., Baratin, A., Arpit, D., Draxler, F., Lin, M., Hamprecht, F.,
  Bengio, Y., Courville, A.: On the spectral bias of neural networks. In:
  International Conference on Machine Learning. pp. 5301--5310. PMLR (2019)

\bibitem{R7}
Ramzi, Z., Chaithya, G., Starck, J.L., Ciuciu, P.: Nc-pdnet: A
  density-compensated unrolled network for 2d and 3d non-cartesian mri
  reconstruction. IEEE Transactions on Medical Imaging  \textbf{41}(7),
  1625--1638 (2022)

\bibitem{TheCascade}
Schlemper, J., Caballero, J., Hajnal, J.V., Price, A., Rueckert, D.: A deep
  cascade of convolutional neural networks for mr image reconstruction. In:
  Information Processing in Medical Imaging: 25th International Conference,
  IPMI 2017, Boone, NC, USA, June 25-30, 2017, Proceedings 25. pp. 647--658.
  Springer (2017)

\bibitem{MLPMIXER}
Tolstikhin, I.O., Houlsby, N., Kolesnikov, A., Beyer, L., Zhai, X.,
  Unterthiner, T., Yung, J., Steiner, A., Keysers, D., Uszkoreit, J., et~al.:
  Mlp-mixer: An all-mlp architecture for vision. Advances in neural information
  processing systems  \textbf{34},  24261--24272 (2021)

\bibitem{ResMLP}
Touvron, H., Bojanowski, P., Caron, M., Cord, M., El-Nouby, A., Grave, E.,
  Izacard, G., Joulin, A., Synnaeve, G., Verbeek, J., et~al.: Resmlp:
  Feedforward networks for image classification with data-efficient training.
  IEEE Transactions on Pattern Analysis and Machine Intelligence  (2022)

\bibitem{MAXIM}
Tu, Z., Talebi, H., Zhang, H., Yang, F., Milanfar, P., Bovik, A., Li, Y.:
  Maxim: Multi-axis mlp for image processing. In: Proceedings of the IEEE/CVF
  Conference on Computer Vision and Pattern Recognition. pp. 5769--5780 (2022)

\bibitem{valanarasu2022unext}
Valanarasu, J.M.J., Patel, V.M.: Unext: Mlp-based rapid medical image
  segmentation network. In: Medical Image Computing and Computer Assisted
  Intervention--MICCAI 2022: 25th International Conference, Singapore,
  September 18--22, 2022, Proceedings, Part V. pp. 23--33. Springer (2022)

\bibitem{DeConvCNN}
Xu, L., Ren, J.S., Liu, C., Jia, J.: Deep convolutional neural network for
  image deconvolution. Advances in neural information processing systems
  \textbf{27} (2014)

\bibitem{MLPMRIDenoise}
Yang, H., Zhang, S., Han, X., Zhao, B., Ren, Y., Sheng, Y., Zhang, X.Y.:
  Denoising of 3d mr images using a voxel-wise hybrid residual mlp-cnn model to
  improve small lesion diagnostic confidence. In: Medical Image Computing and
  Computer Assisted Intervention--MICCAI 2022: 25th International Conference,
  Singapore, September 18--22, 2022, Proceedings, Part III. pp. 292--302.
  Springer (2022)

\bibitem{ye2022multi}
Ye, Y., Lyu, J., Hu, Y., Zhang, Z., Xu, J., Zhang, W.: Multi-parametric mr
  imaging with flexible design (multiplex). Magnetic Resonance in Medicine
  \textbf{87}(2),  658--673 (2022)

\bibitem{R4}
Zhang, H., Shinomiya, Y., Yoshida, S.: 3d mri reconstruction based on 2d
  generative adversarial network super-resolution. Sensors  \textbf{21}(9),
  ~2978 (2021)

\bibitem{AUTOMAP}
Zhu, B., Liu, J.Z., Cauley, S.F., Rosen, B.R., Rosen, M.S.: Image
  reconstruction by domain-transform manifold learning. Nature
  \textbf{555}(7697),  487--492 (2018)

\end{thebibliography}

\end{document}